\documentclass[11pt,twoside]{article}

\usepackage{asp2006}
\usepackage{epsf}
\usepackage{psfig}
\usepackage{lscape}
\usepackage{graphicx}

\newcommand{\etal}{{et al.~}}

\begin{document}
\title{
Frequency and Impact of Galaxy Mergers and Interactions over the last 7 Gyr
}

 \author{
  S.  Jogee \altaffilmark{1},  
   S. Miller \altaffilmark{1},  
  K. Penner \altaffilmark{1}, 
  E. F. Bell \altaffilmark{2}, 
  C. Conselice  \altaffilmark{3}, 
  R. E. Skelton \altaffilmark{2}, 
  R. Somerville \altaffilmark{2},
  H-W. Rix  \altaffilmark{2}, 
   F. D. Barazza  \altaffilmark{4},
   M. Barden  \altaffilmark{5},
   A. Borch  \altaffilmark{2}, 
   S. V. Beckwith \altaffilmark{6}, 
   J. A.  Caldwell \altaffilmark{7}, 
   B.  H{\"a}u{\ss}ler  \altaffilmark{3}, 
   C. Heymans  \altaffilmark{8,15}, 
   K. Jahnke \altaffilmark{2},
   D. McIntosh \altaffilmark{9}, 
   K. Meisenheimer  \altaffilmark{2},
   C. Papovich  \altaffilmark{10},  
   C. Peng \altaffilmark{11},  
   A. Robaina  \altaffilmark{2},
   S. Sanchez  \altaffilmark{12},  
   L. Wisotzki  \altaffilmark{13},  
   C. Wolf  \altaffilmark{14}
%  X. Z. Zheng (PMO)\altaffilmark{16}
}

  \altaffiltext{1}
{University of Texas at Austin,  Austin, TX 78712-0259}
   \altaffiltext{2}
 {Max-Planck-Institut f\"{u}r Astronomie, Heidelberg, Germany}
   \altaffiltext{3}
  {The University of Nottingham, Nottingham NG7 2RD, UK}
  \altaffiltext{4}
 {Laboratoire d'Astrophysique, EPFL, 1290 Sauverny, Switzerland}
  \altaffiltext{5}
  {University of Innsbruck, A-6020 Innsbruck,   Austria}
  \altaffiltext{6}
 {Johns Hopkins University,   Baltimore, MD 21218}
  \altaffiltext{7}
  {University of Texas, McDonald Observatory, Fort Davis TX, 79734 USA}
  \altaffiltext{8}
   {University of British Columbia, Vancouver, V6T 1Z1, Canada}
 \altaffiltext{9}
  {University of Massachusetts, Amherst, MA 01003, USA}
  \altaffiltext{10}
 {University of Arizona, Steward Observatory, Tucson, AZ 85721}
    \altaffiltext{11}
    {NRC Herzberg Institute of Astrophysics, Victoria, Canada}
    \altaffiltext{12}
    {Centro Astronómico Hispano Alemán, Calar Alto, E-04004 Almeria, Spain}
   \altaffiltext{13}
   {Astrophysikalisches Institut Potsdam, D-14482 Potsdam, 
    Germany}
  \altaffiltext{14}
  {University of Oxford, Keble Road, Oxford OX1 3RH, U.K.}
  \altaffiltext{15}
  {Institut d'Astrophysique de Paris, 75014 Paris, France}
%  \altaffiltext{16}
%{Purple Mountain Observatory, Nanjing, China.}

%%%%%%%%%%%%%%%%%%%%%%%%

\begin{abstract} %%% Abstract to run on from here.
We explore the history and impact of galaxy mergers and interactions 
over  $z \sim$~0.24--0.80, based on $HST$ ACS, Combo-17, 
and Spitzer 24~$\mu$m data of $\sim$~4500 galaxies in the GEMS survey.
Using visual and quantitative parameters,
we identify galaxies with strong distortions indicative of recent 
strong interactions and mergers versus normal galaxies (E/S0, Sa, Sb-Sc, 
Sd/Irr). 
%with no 
%significant externally-triggered asymmetries 
Our preliminary results are: 
(1)~ The observed fraction $F$ of strongly disturbed systems  among 
high mass  ($M \ge$~$2.5 \times 10^{10}$ $M_{\odot}$) 
galaxies is $\sim$~9\% to 12\% in  every Gyr bin over  
$z\sim$~0.24--0.80. The corresponding merger rate is $\sim$~ a 
few $\times 10^{-4}$ galaxies Gyr$^{-1}$ Mpc$^{-3}$. 
The fraction $F$ shows fair agreement with the merger 
fraction of mass ratio $\ge$~1:10 predicted by several 
LCDM-based simulations.
(2)~ For $M \ge$~$1.0 \times 10^{9}$ $M_{\odot}$  systems,  
the average SFR  of strongly disturbed systems is only 
modestly enhanced with respect to normal galaxies.
In fact, over $z \sim$~0.24--0.8,  strongly disturbed systems only 
account for a small fraction ($<$ 30\%)  of the total SFR density.
This suggest that the behaviour  of the cosmic SFR density 
over $z \sim$~0.24--0.80 is  predominantly shaped by normal galaxies.
%(4) The specific SFR is on average higher at lower masses, 
%suggesting that at $z \le$~0.80, star formation produces a 
%larger fractional growth in stellar  mass in lower mass systems.
\end{abstract}

%%% MAIN BODY OF TEXT GOES HERE. CONSULT "INSTRUCTIONS FOR AUTHORS USING
%%% LATEX2E MARKUP", SECTIONS 2.3-2.6 FOR HELP WITH EQUATIONS, FIGURES,
%%% AND TABLES.

%\section{}   %%% Top level section head (remove "%" symbol)
%\subsection{}   %%% Second level section head (remove "%" symbol)
%\subsubsection{}   %%% Lowest level section head (remove "%" symbol)
%\section*{}    %%% Unnumbered top level section head (remove "%" symbol)
%\subsection*{}   %%% Unnumbered second level section head (remove "%" symbol)

\vspace{-0.8 cm}
\section{Introduction}

%\vspace{-0.2 cm}
%\noindent
Hierarchical  $\Lambda$CDM  (HLCDM)  models provide a successful paradigm 
for the growth of dark matter on large scales, but predictions of how 
galaxies evolve depend on the baryonic merger 
history, star formation,  feedback, and other aspects of the baryonic 
physics. In this paper, we provide empirical constraints on the frequency
of mergers and interactions, and their impact on the star formation (SF) 
of galaxies over the last seven Gyr (Jogee \etal 2008).

Our sample consists of $\sim$~4500 galaxies with $R_{\rm Vega} \le$ 24 
over the redshift interval $z\sim$~0.24--0.80 
(lookback $T_{\rm back}$~$\sim$~3--7 Gyr\footnote{We assume in this
paper a flat cosmology with $\Omega_M = 1 - \Omega_{\Lambda} = 0.3$
and $H_{\rm 0}$ =70~km~s$^{-1}$~Mpc$^{-1}$.}), drawn from the 
Galaxy Evolution from Morphology and SEDS (GEMS; Rix \etal 2004) survey.
We use  {\it HST} ACS F606W ($V$) images  from GEMS, accurate 
spectrophotometric redshifts  [$\delta_{\rm z}$/(1 + $z$)~$\sim$~0.02 
down to  $R_{\rm Vega}$ = 24] from COMBO-17 (Wolf \etal 2004), 
stellar masses  based on COMBO-17 (Borch \etal 2006), and 
star formation rates (Bell \etal 2007)  based on Combo-17 UV and Spitzer 
%24~$\mu$m  
% (Rieke \etal 2004; Papovich \etal 2004) 
data.

\vspace{-0.4 cm}
\section{Analysis and Preliminary Results }

\vspace{-0.2 cm}
%\noindent
We visually classified galaxies in the F606W images into the 
following two main visual classes (VCs).
%, with the main goal of identifying 
%galaxies with  recent strong interactions and mergers:
(1) `Dist' : This VC contains strongly disturbed galaxies with
{\it externally-triggered} distortions  indicative of a
recent strong tidal  interaction or merger.
% with a significant mass ratio (>1:10)
The distortions include arcs, shells, ripples, tidal debris,  
warps, offset rings,  extremely asymmetric light distributions, tidal 
tails, and  multiple nuclei inside a common body. 
(2) `Normal': This VC contains the remaining galaxies. 
It includes  E/S0, Sa, and Sb-Sd galaxies with no significant 
{\it externally-triggered} distortions, and the class ``Irr1'' of 
galaxies with {\it internally-triggered} asymmetries described
below. Many non-interacting galaxies have some low level of small-scale 
asymmetry in their rest-frame $V$ or $B$ band light due to star-forming 
regions  or due to the low ratio of rotational to random velocities
in  low mass systems   (e.g., in local  Im and 
Sm galaxies).  These {\rm internally-triggered} asymmetries 
differ in scale (few 100 pc vs several kpc) and morphology 
from the {\rm externally-triggered} distortions listed in (1).

We also ran the  CAS code (Conselice et al. 2000) on the the F606W 
images to derive  asymmetry (A), and clumpiness (S) parameters.
The CAS  criterion  ($A >$~0.35 and $A > S$)  picks up only  37\% to 58\% 
of the  strongly disturbed  'Dist' galaxies, as expected from simulations
(Conselice 2006). However, it is also contaminated by a significant 
number of relatively normal galaxies.

The visual classification was performed by three classifiers to measure
the dispersion. The largest uncertainties in the VCs 
are in the last bin  ($z\sim$~0.6 to 0.8) due to surface brightness 
dimming and due to the  rest frame wavelength ($\lambda_{\rm rest}$) 
shifting to the violet/near-UV  (3700 \AA to 3290 \AA). We addressed 
these by verifying that the VCs of  most 
strongly disturbed and normal galaxies do not change when we repeat 
the classification 
in all 4 bins using the deep GOODS F850LP images, whose  redder 
pivot wavelength  (9103 \AA),  
ensures that $\lambda_{\rm rest}$~$\ge$~5000 \AA \ in 
all bins.  Further tests, including  Monte Carlo simulations and 
comparisons with HUDF images are under way.
We summarize our current results below:

\vspace{0.1 cm}
{\it 1.}
Fig.~1 shows the color-mass distribution of the sample.
Out to $z\sim$~0.8, the req sequence is complete for  high  mass 
($M\ge$~$2.5 \times 10^{10}$ $M_{\odot}$; $N$=804) galaxies 
(Borch \etal 2006), while the blue cloud is complete for intermediate 
mass ($M\ge$~$1.0 \times 10^{9}$ $M_{\odot}$; $N$=3864) systems.  
The strongly disturbed (Dist) galaxies, coded as orange stars, lie
on both the red sequence and blue cloud.  

\vspace{0.05 cm}
{\it 2.}
The observed fraction $F$ of strongly disturbed systems  among 
high mass  ($M \ge$~$2.5 \times 10^{10}$ $M_{\odot}$) 
galaxies is $\sim$~8\% to 12\% in each  Gyr bin over  $z \sim$~
0.24--0.80 (Fig.~2). Similar results are  reported by Lotz \etal
(2008). The corresponding merger rate is $\sim$~ a 
few $\times 10^{-4}$ galaxies Gyr$^{-1}$ Mpc$^{-3}$. 

\vspace{0.05 cm}
{\it 3.}
Fig.~2 compares the observed  fraction $F$ with the 
fraction of  major  mergers (M1/M2~$\ge$~1:4; dashed lines) and 
major+minor mergers (M1/M2~$\ge$~1:10; solid lines)
predicted  by  LCDM-based semi-analytical (SA), N-body, and SPH 
simulations, namely 
Hopkins \etal (2008; red lines labeled `H'; SA),  Somerville \etal (in prep.; 
blue lines labeled `S'; SA), Benson \etal (2005; pink lines labeled `B'; SA),  
% ;Maller \etal (2006; green line labeled `M'; SPH), 
and  D'Onghia \etal (2008; labeled `D'; N-body). 
There is fair agreement between
$F$ and the predicted fraction of mergers with M1/M2~$\ge$~1:10.

%  \vspace{0.0 cm}
%  \\noindent 
%  \{\it 4.}
%  \The  UV-based SFR  ranges from  $\sim$~0.01 to 25 $M_{\odot}$ yr$^{-1}$,
%  \with most galaxies having a rate below 5  $M_{\odot}$ yr$^{-1}$.
%  \For the galaxies with both UV and IR data, the  median value of 
%  \(SFR$_{\rm IR}$/SFR$_{\rm UV}$) is $\sim$~3.6, indicating a 
%  substantial  amount of obscured star formation. 

\vspace{0.05 cm}
%\noindent 
{\it 4.}
Among  $M \ge$~$1.0 \times 10^{9}$ $M_{\odot}$  systems 
over  $z \sim$~0.24--0.80, the average SFR of strongly disturbed
systems is only modestly enhanced with respect to normal galaxies 
over this redshift range (Fig.~3; see also Robaina et al. in prep.). 
This modest enhancement is consistent with a recent statistical studies 
of the SF efficiency from based on large sets of numerical simulations 
(di Matteo et al 2007).
In  fact, for this mass range, strongly disturbed systems only 
account for a small fraction ($<$ 30\%) of the cosmic SFR density over
 $z\sim$~0.24--0.80 (Fig.~4).  These results complement the findings 
that normal galaxies dominate the UV  (Wolf \etal 2005) and IR (Bell 
\etal 2005) luminosity density  at $z \sim $~0.65--0.75, and 
are in good agreement  with the predictions by Hopkins \etal (2006).
Our results  suggest that 
the corresponding behaviour  of the cosmic SFR density 
over $z \sim$~0.24--0.80 is  predominantly shaped by normal galaxies.
%Furthermore, the specific SFR is on average higher at lower masses, 
%suggesting that at $z <$~0.80,  SF  produces a  larger 
%fractional growth in the stellar mass of lower mass systems.

\vspace{-0.05 cm}
\acknowledgements %%% Text of acknowledgements runs on after this command.
S.J. acknowledge support from NSF grant AST 06-07748, NASA LTSA grant 
NAG5-13063, and HST G0-11082 from STScI
%, which is operated by AURA, Inc., for NASA, under NAS5-26555.

%%%%%   Makes 2 x2 panel using  \includegraphics without \begin{figure}
% The \includegraphics gives more flexibility in y and x-positioning
% of all 4 figures. If use \includegraphics without  \begin{figure} then 
%  - can make up tailored caption to call the 4 figures as Fig 1, Fig 2
%    Fig 3, Fig 4 

\clearpage
\vskip -7.0 mm      %  the \vskip seems to have no effect at top of page
\vspace{-7.0 mm}    %  the \vspace seems to have no effect at top of page
\hspace{-2.0 cm}
\includegraphics[width=2.9 in]{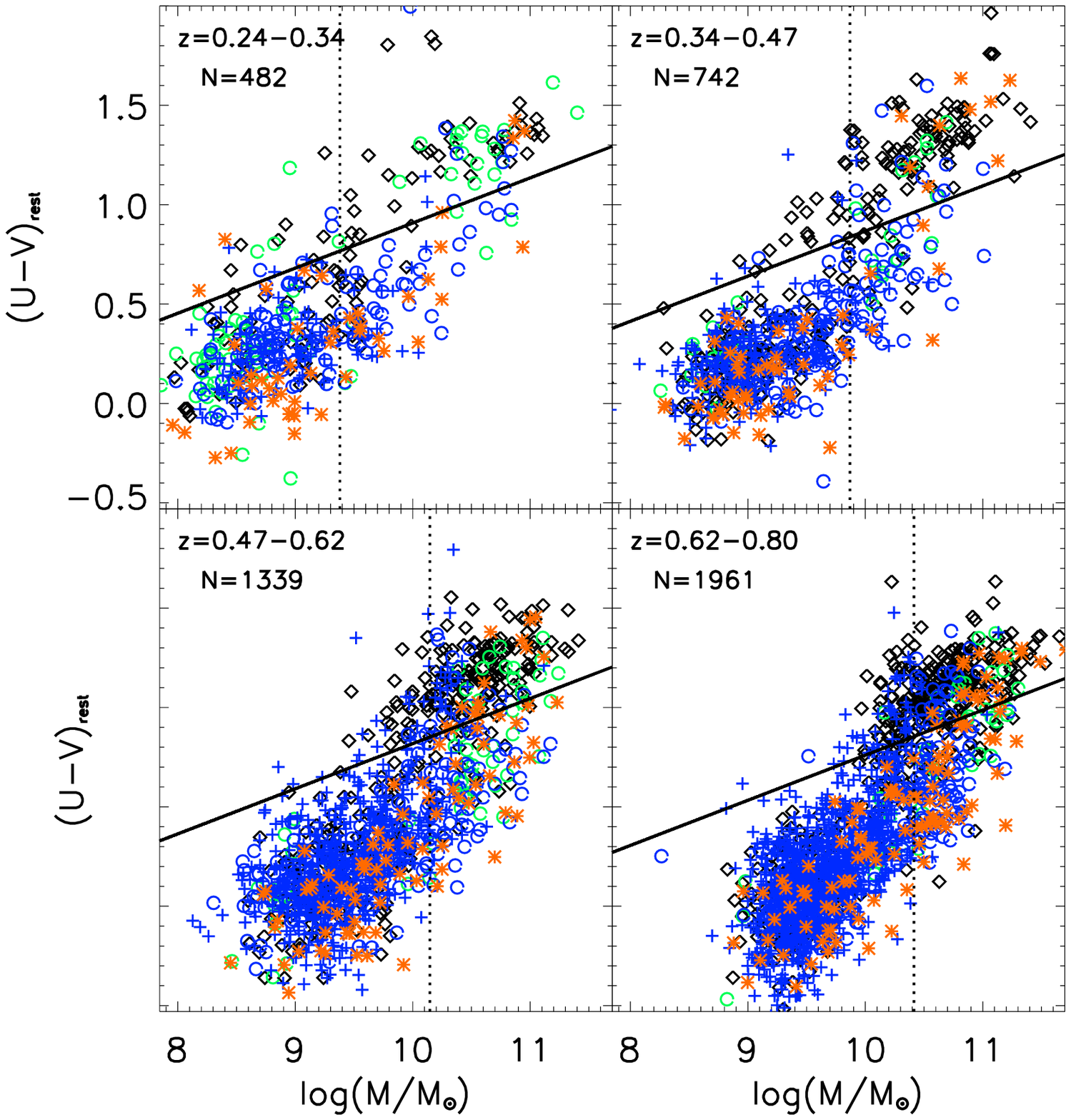}   %%%%%{plt.colms.f606.4panel.eps}
\vspace {0.5 mm}
\hspace{2.0 mm}
\includegraphics[width=2.6in]{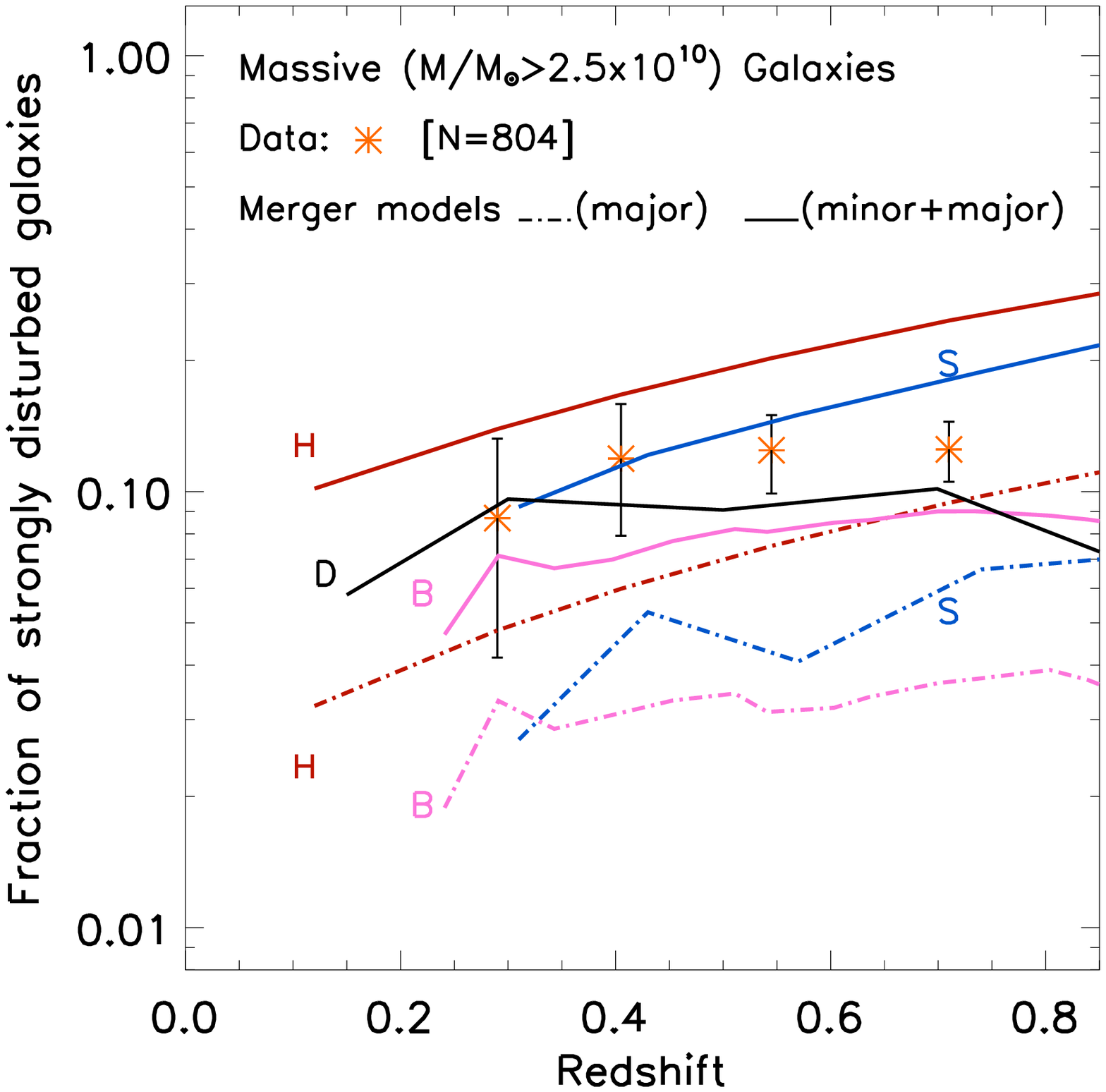}    %%%% {plot-merge-mass.plt71.eps}
\vskip 0.3 cm    % Use \vskip rather than \\ or \vspace{0 mm} for line break!! 
\vspace{0.0 mm}   
\hspace{-1.6 cm}
\includegraphics[width=2.63 in]{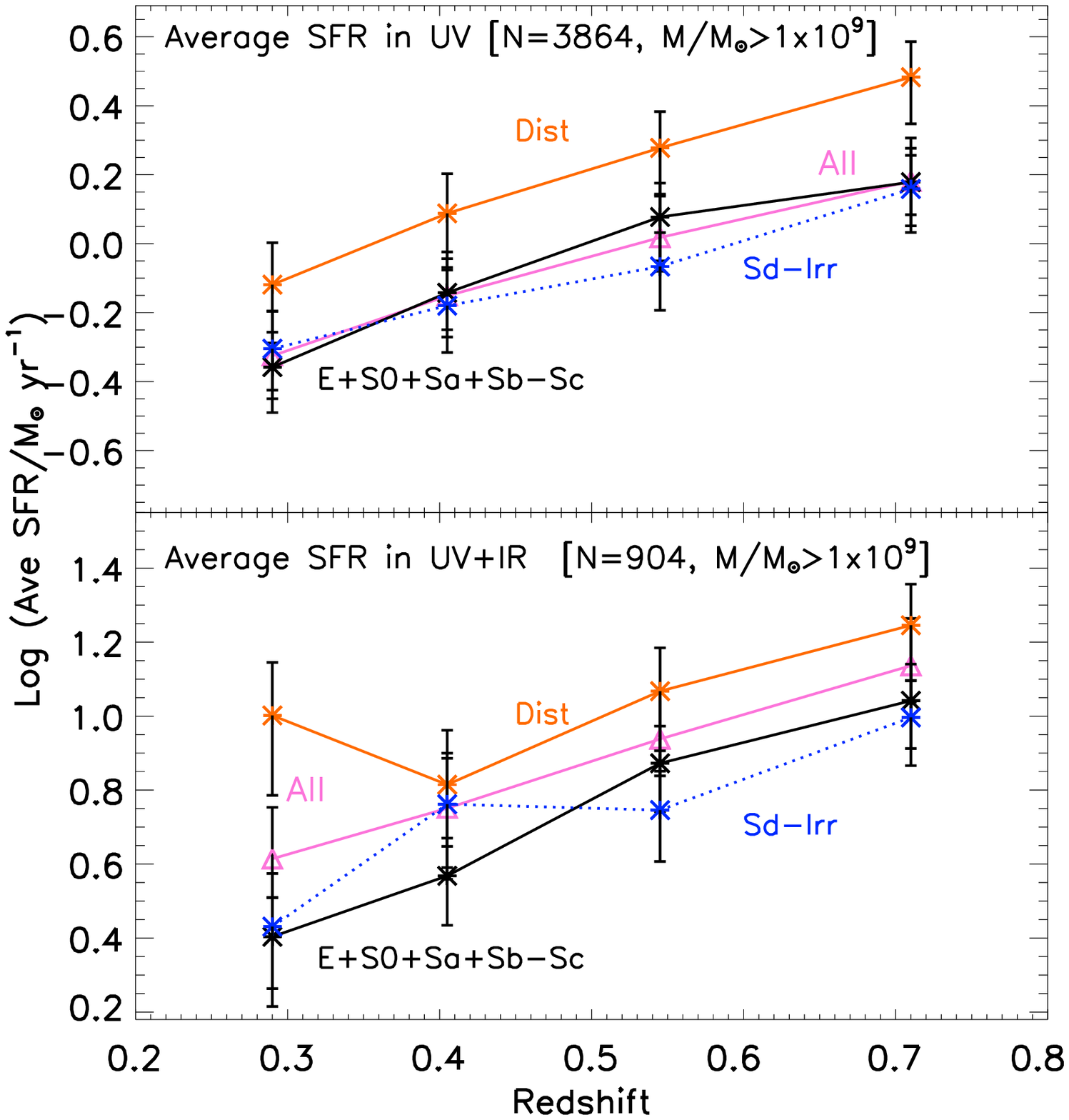}   %%%%% {plot.sfr.plt2new.eps}
\hspace{5.0 mm}
\includegraphics[width=2.73 in]{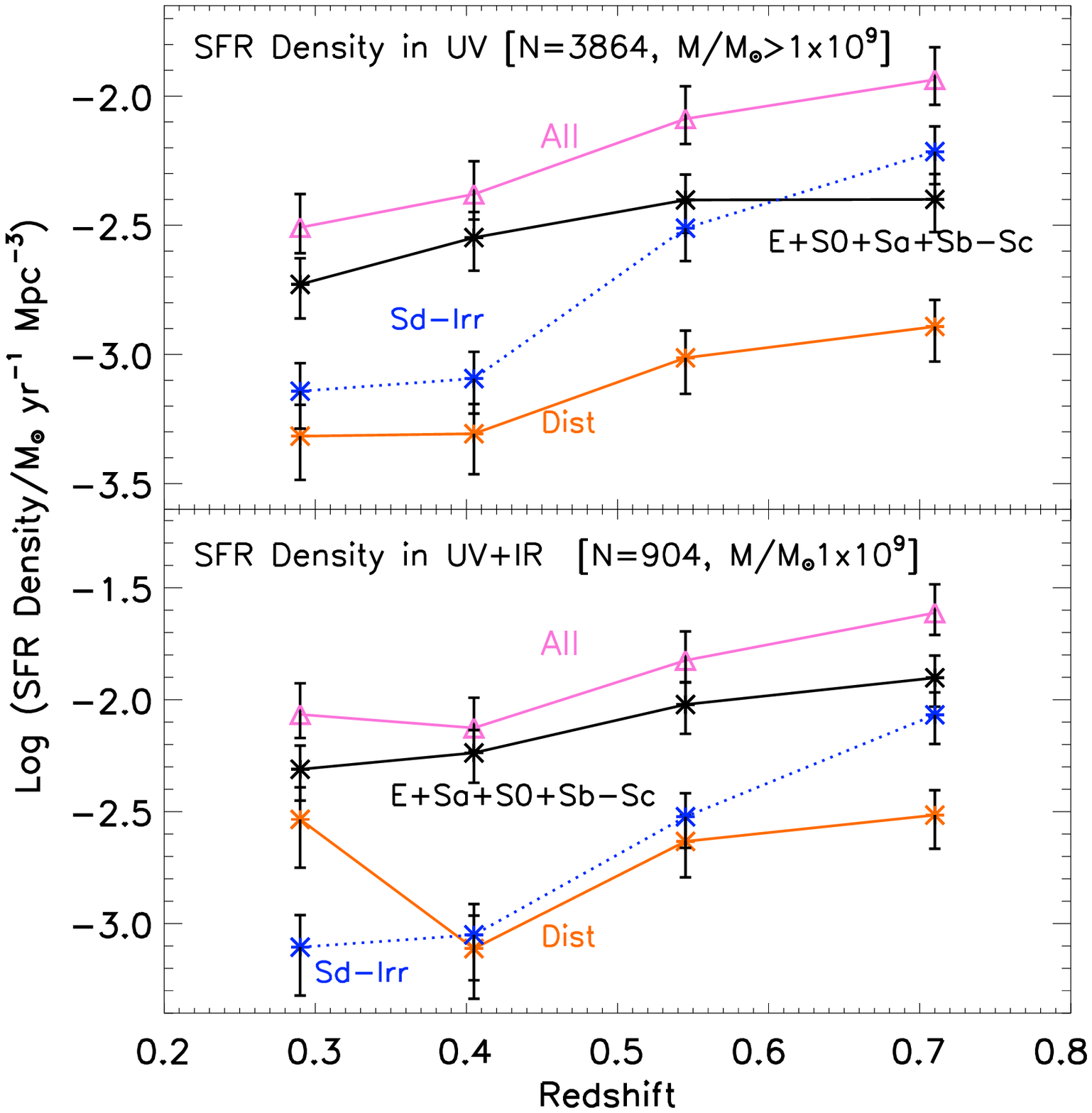}   %%%%%{plot.sfr.plt1new.eps}
\\
\vskip 0 mm    % Use \vskip rather than \\ or \vspace{0 mm} for line break!!
\noindent 
{\bf Fig.~1~(Top Left):}
The rest-frame $U-V$ color is plotted {\it vs} the stellar mass in 
four redshift bins, which span 1 Gyr each.
%, and cumulatively cover  
%the interval $z \sim$~0.24--0.80 ($T_{\rm back}$~$\sim$~3--7 Gyr). 
%$N$ denotes the number of galaxies in each bin. 
The diagonal line  marks the red sequence, 
% at the average redshift $z_{\rm ave}$ of the bin. 
and the vertical line denotes its mass completeness limit.
Blue cloud galaxies are complete well below this mass.  
Strongly disturbed systems are coded as orange stars.
%and normal relatively undisturbed  as black diamonds (E+S0), green
% circles (Sa),  blue circles (Sb-Sc) and blue crosses (Sd-Irr).
{\bf Fig.~2~(Top Right):}
The observed  fraction  of strongly disturbed
galaxies  (orange stars)  is compared to predictions from 
 theoretical HLCDM-based simulations. See text for details. 
{\bf Fig. 3~(Lower Left):}
The average SFR  
based on UV data and  UV+IR data (for galaxies with a  
24~$\mu$m detection) is plotted for strongly disturbed and 
normal galaxies. 
% See text for details.
%  and strongly distorted galaxies   
% The average SFR of strong disturbed/interacting  (Dist/Int) systems is
% only enhanced by a small factor of 2-3 compared to normal  relatively 
% undisturbed galaxies.
{\bf Fig.~4~(Lower Right):}
The contribution of strongly disturbed galaxies and normal 
galaxies to  the cosmic  SFR density  out to $z \sim$~0.80 
($T_{\rm back}$~$\sim$~7 Gyr) is shown.
%The former  only contribute a small fraction (typically 20\%) of the 
%UV-based (top panel) or total (lower panel) cosmic SFR density.


\vspace{-0.2 cm}
\begin{thebibliography}{}

\vspace{-0.1 cm}
\bibitem[]{} Bell, E.~F., et al.\ 2005, \apj, 625, 23 
\bibitem[]{} Bell, E.~F., et al. 2007, \apj, 663, 834
\bibitem[]{} Benson, A.~J., Kamionkowski, M., \& Hassani, S.~H.\ 2005, \mnras, 357, 847 
\bibitem[]{} Borch, A., et al. 2006, \aap, 453, 869
\bibitem[]{} Conselice, C.,\ Bershady, M. A.,\ \&  Jangren, A. 2000, \apj, 529, 886
\bibitem[]{} Conselice, C.~J.\ 2006,\apj, 638, 686
\bibitem[]{} di Matteo, P., Combes, F., Melchior, A.-L., \& Semelin, B.\ 2007, \aap, 468, 61 
\bibitem[]{} D'Onghia, E., Mapelli, M. , Moore, B. 2008, MNRAS,  submitted
\bibitem[]{} Jogee, S.  et al.  2008, ApJ, in preparation
\bibitem[]{} Hopkins, P.~F., Somerville, R.~S., Hernquist, L., Cox, T.~J., Robertson, B., \& Li, Y.\ 2006, \apj, 652, 864 
\bibitem[]{} Hopkins \etal  2008, \apj, submitted (arXiv:0706.1243)
\bibitem[]{} Lotz, J.~M., et al.\ 2008, \apj, 672, 177 
%\bibitem[]{} Maller, A.~H \etal  2006, \apj, 647, 763
%\bibitem[]{} Maller, A.~H., Katz, N., Kere{\v s}, D., Dav{\'e}, R., \& Weinberg, D.~H.\ 2006, \apj, 647, 763
\bibitem[]{} Rix, H.-W., et al.\ 2004, \apjs, 152, 163
%\bibitem[]{} Rieke \etal 2004
\bibitem[]{} Wolf, C., et al.\ 2004, \aap, 421, 913
\bibitem[]{} Wolf, C., et al.\ 2005, \apj, 630, 771 


%\bibitem[Wolf et al. 2003]{w03} Wolf, C., \ Meisenheimer, K., Rix
%  H.-W., \ Borche, A., Kleinheinrich, M., \ 2003, \aa, 401, 73
\end{thebibliography}
\end{document}